\documentclass{iopart}

\usepackage[dvips]{graphicx}
\usepackage{iopams}

\begin{document}

\noindent
{\small Proc. of GWDAW-10}
\hfill{\small LIGO-P060005}

\title[Tikhonov regularization for gravitational wave bursts]
{Rank deficiency and Tikhonov regularization in the inverse 
problem for gravitational-wave bursts}

\author{M~Rakhmanov
\footnote{Current address: LIGO Hanford Observatory, PO Box 159,
Richland, WA 99352, USA}}

\address{Center for Gravitational Wave Physics, The Pennsylvania State 
University, 104 Davey Laboratory, University Park, PA 16802, USA}

\ead{malik@ligo-wa.caltech.edu}

\begin{abstract}
Coherent techniques for searches of gravitational-wave bursts
effectively combine data from several detectors, taking into account
differences in their responses. The efforts are now focused on the
maximum likelihood principle as the most natural way to combine data,
which can also be used without prior knowledge of the signal. Recent
studies however have shown that straightforward application of the
maximum likelihood method to gravitational waves with unknown
waveforms can lead to inconsistencies and unphysical results such as
discontinuity in the residual functional, or divergence of the
variance of the estimated waveforms for some locations in the sky. So
far the solutions to these problems have been based on rather different
physical arguments. Following these investigations, we now find that 
all these inconsistencies stem from rank deficiency of the underlying 
network response matrix. In this paper we show that the detection of 
gravitational-wave bursts with a network of interferometers belongs 
to the category of ill-posed problems. We then apply the method of 
Tikhonov regularization to resolve the rank deficiency and introduce 
a minimal regulator which yields a well-conditioned solution to the 
inverse problem for all locations on the sky.
\end{abstract}


\section{Introduction}

Efforts in searches for bursts with gravitational-wave detectors are
now devoted to the development and testing of coherent data analysis
techniques which do not require prior knowledge of the signal. Several
such methods have been proposed in the past, beginning with the
method of G\"{u}rsel and Tinto \cite{Gursel:1989} which substantially
predates all other techniques. The approach of G\"{u}rsel and Tinto 
is based upon explicit construction of a null data stream for
two-polarization gravitational waves with completely arbitrary
waveforms. Renewed interest in this method is motivated by the 
search for efficient vetoes of bursts of non-astrophysical origin
\cite{Wen:2005}. Other methods which do not rely on the waveforms
include coherent power filters \cite{Sylvestre:2003} and
cross-correlations of data from different detectors 
\cite{Cadonati:2004, Cadonati:2005, Rakhmanov:2005}. Recent studies 
are however converging on the maximum likelihood method as the most 
natural way to combine data from a network of gravitational wave
detectors. It was shown by Flanagan and Hughes that the maximum
likelihood inference can, in principle, be used without any knowledge
of the anticipated signal \cite{Flanagan:1998b}. This observation led
to the development of a data analysis technique known as excess power 
\cite{Anderson:2001}. Subsequent extensions of this approach included
Karhunen-Loeve expansions \cite{Vicere:2002} and non-parametric
adaptive filters \cite{Chassande-Mottin:2003}. This groundwork helped
building our confidence in the maximum likelihood method as a
general framework for searches of gravitational-wave bursts with
unknown waveforms. However, it was recently discovered that
straightforward application of the maximum likelihood principle can
lead to inconsistencies and unphysical results such as discontinuity
in the residual functional called the two-detector paradox
\cite{Johnston:2004, Klimenko:2005}, or unphysically large variations 
in the estimated signal-to-noise ratio \cite{Mohanty:2006}. The
proposed solutions included constraints and penalty functions
derived from various physical arguments \cite{Klimenko:2005, 
Mohanty:2006}. Continuing these investigations, we now consider the
problem from a very general point of view. In this paper, we show that
all these inconsistencies and paradoxes arise because the inverse
problem for bursts belongs to the category of ill-posed discrete
(matrix) problems. In particular, its underlying matrix of
coefficients suffers from rank deficiency. It is then natural to look
for a solution within the Tikhonov regularization approach
\cite{Tikhonov:1977} which is the general framework for solving
ill-posed problems in mathematical physics.

\section{The inverse problem for bursts}
\subsection{Network response}

\begin{figure}[t]
   \includegraphics[width=\textwidth]{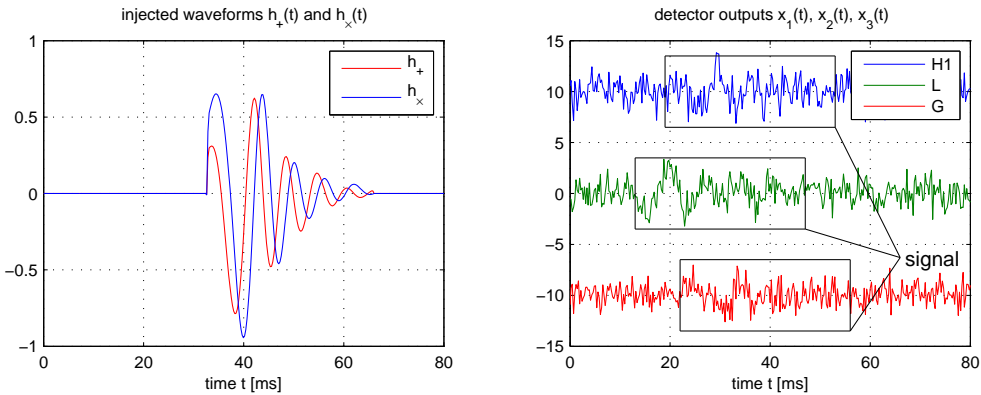}
   \caption{Simulated waveform injection. {\it Left:} a typical
   two-polarization waveform from numerical modeling of binary black
   hole coalescence \cite{Baker:2002}. {\it Right:} modeled detector
   outputs in H1-L-G network for a source located at 
   $\phi = -60^{\circ}$ and $\theta = 22^{\circ}$,
   with the matched-filter SNR \cite{Cadonati:2004} of 8, 12 and 7.}
   \label{signalsHLG}
\end{figure}

Consider a network of $m$ detectors located at different places on 
Earth. The response of the detectors to gravitational waves 
with two polarizations, $h_{+}$ and $h_{\times}$, is given by
\begin{equation}\label{xi=FhFh}
   \xi_i(t) = 
      F_{i+}(\phi,\theta) \,  h_{+}(t) + 
      F_{i\times}(\phi,\theta) \, h_{\times}(t) ,
\end{equation}
where $F_{i+}$ and $F_{i\times}$ are the antenna-pattern functions, and
$\phi$ and $\theta$ are the spherical angles of the source in the sky. 
In general, the data from the network contains both signal and noise
(see figure \ref{signalsHLG}):
\begin{equation}\label{sigNoise}
   x_i(t) = \xi_i(t + \tau_i) + \eta_i(t) ,
\end{equation}
where the noise terms $\eta_i$ are assumed to be statistically 
independent among the detectors.
The delays $\tau_i = \tau_i(\phi,\theta)$ depend on the source
location and are calculated with respect to a common reference, 
usually taken at the center of Earth. After changing the
variables: $t \rightarrow t - \tau_i$, we can write 
(\ref{xi=FhFh}-\ref{sigNoise}) in the matrix form:
\begin{equation}\label{x=TAh}
   x(t|\phi,\theta) = A(\phi,\theta) \, h(t) + \eta(t|\phi,\theta) ,
\end{equation}
where $x$, $h$, and $\eta$ are column vectors and $A$ is 
$m\! \times \!2$ matrix,
\begin{equation}\label{matrixA}
   A = \left[
   \begin{array}{cc}
      F_{1 +}  &  F_{1 \times} \\
      \vdots   &  \vdots   \\
      F_{m +}  &  F_{m \times}
   \end{array}\right] ,
\end{equation}
which will be called the network response matrix. The notation 
$x(t|\phi,\theta)$ implies that each component of the vector 
is shifted by its appropriate time delay:
\begin{equation}
   x_i(t|\phi,\theta) = x_i[t - \tau_i(\phi,\theta)] .
\end{equation}
For simplicity, we will often omit $(\phi,\theta)$ and write
(\ref{x=TAh}) as
\begin{equation}
   x(t) = A \, h(t) + \eta(t) .
\end{equation}
Also, it will be convenient to view the matrix $A$ as comprised of two
vectors: 
\begin{equation}
   A = [ \, {\mathbf{F}}_{+} \; {\mathbf{F}}_{\times} ] ,
\end{equation}
known as the {\it range vectors} of $A$ \cite{Golub:1989}. Examples 
of numerical calculations in this paper correspond to LIGO Hanford 
(H1 and H2), LIGO Livingston (L), VIRGO (V), and GEO-600 (G) detectors.

\subsection{Moore-Penrose inverse}

The inverse problem can be formulated as follows: given data from $m$ 
detectors $x_i(t)$ find the gravitational-wave amplitudes $h_i(t)$ and 
the source location in the sky $(\phi, \theta)$ by solving 
\begin{equation}
   A(\phi, \theta) \, h(t) = x(t| \phi, \theta) ,
\end{equation}
in which the data is contaminated with noise. In general, the problem 
cannot be solved exactly, and one looks for an approximate solution 
by minimizing the functional
\begin{equation}\label{LSQdef}
   L[h] = \| x(t) - A \, h(t) \|^2 ,
\end{equation}
where $\| . \|$ stands for vector 2-norm:
\begin{equation}
   \| x(t) \|^2 = \sum_{i=1}^{m} \int_0^T x_i^2(t) \, dt ,
\end{equation}
defined over a suitable interval of observation $T$. The least squares 
(LSQ) functional $L[h]$ is usually derived within the maximum
likelihood approach.

Variation of the functional $L[h]$ with respect to $h_i(t)$ yields the 
{\it normal equations}, which in the matrix form can be written as
\begin{equation}
   M \, h(t) = A^T x(t) , \qquad {\mathrm{where}} \qquad M = A^T A .
\end{equation}
Then the solution is given by
\begin{equation}\label{LSQsolution}
   h(t) = A^{\dagger} \, x(t) ,
\end{equation}
where the $2\! \times \!m$ matrix 
\begin{equation}\label{MPinverse}
   A^{\dagger} = M^{-1} A^T 
\end{equation}
is known as the Moore-Penrose inverse or {\it pseudoinverse} of 
matrix $A$ \cite{Greville:1959}.

\begin{figure}[t]
   \includegraphics[width=\textwidth]{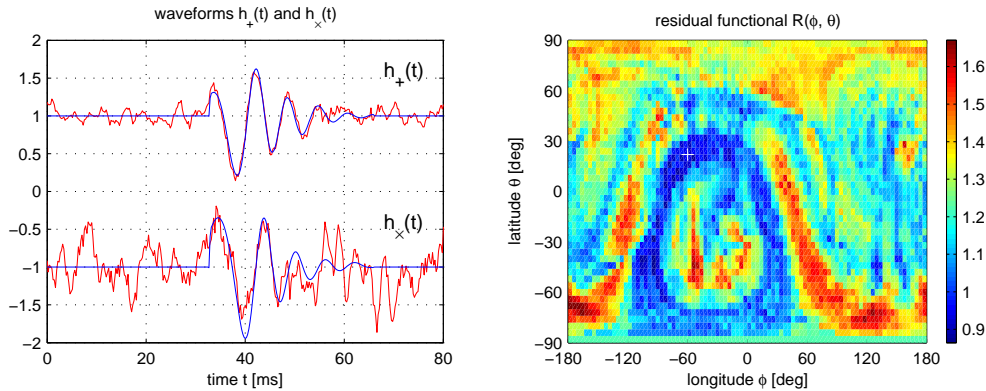}
   \caption{Typical search results for the simulated waveform injection 
   shown in figure~\ref{signalsHLG}. {\it Left:} injected (blue) and 
   estimated (red) waveforms (after low-pass filtering). {\it Right:} 
   sky map of the residual functional (\ref{defResidual}). The
   position of the source $(-60^{\circ}, 22^{\circ})$ is marked by $+$.}
   \label{resultsHLG}
\end{figure}

Consider now a situation when the data contains a signal,
\begin{equation}\label{dataSignal}
   x(t) = A(\phi_s, \theta_s) h_s(t) + \eta(t) ,
\end{equation}
where $(\phi_s, \theta_s)$ represent the source position in the sky.
Then the solution acquires a non-zero expectation value:
\begin{equation}
   \langle h(t) \rangle = 
      A^{\dagger}(\phi, \theta) A(\phi_s, \theta_s) \, h_s(t) .
\end{equation}
At the true location of the source $\langle h(t) \rangle = h_s(t)$,
i.e. the solution given by the Moore-Penrose inverse yields an
un-biased estimation of the gravitational-wave amplitudes. An example
of a solution for $h(t)$ obtained from noisy data is shown in 
figure~\ref{resultsHLG}. Note that (\ref{LSQsolution}) provides the
solution to the first part of the inverse problem: determination of
$h(t)$. In the next section we describe the solution to the second
part of the problem: determination of the source position on the sky
$(\phi, \theta)$.

Introduction of the Moore-Penrose inverse allows us to define two 
complementary and orthogonal projector matrices:
\begin{equation}\label{defPQ}
   P = A A^{\dagger}, 
      \qquad {\mathrm{and}} \qquad 
   Q = I - A A^{\dagger} .
\end{equation}
Note that $P A = A$ and $Q A = 0$ which means that $P$ projects onto the
vector space of the {\it range} of matrix $A$ and $Q$ projects onto
the {\it null space} of $A^T$, which is a subspace complementary to
the range space.

\subsection{The residual functional}

Substitution of the solution $h(t)$ (\ref{LSQsolution}) 
into $L[h]$ (\ref{LSQdef}) yields the residual functional:
\begin{equation}\label{defResidual}
   R = \| Q x \|^2 = x^T Q \, x ,
\end{equation}
where we used the fact that $Q^2 = Q$. This functional will be used to
find the location of the source in the sky. We will now prove that the 
expectation value of $R(\phi, \theta)$ reaches minimum at the true
location of the source. Assume that the data contains a signal
(\ref{dataSignal}), then the residual functional is given by
\begin{equation}\label{resSigNoise}
   R = \| Q A_s h_s \|^2 + 2 \eta^T Q A_s h_s  + \eta^T Q \, \eta ,
\end{equation}
where $A_s \equiv A(\phi_s, \theta_s)$. The first term in
(\ref{resSigNoise}) contains only deterministic quantities and
therefore its expectation value is the same as its current value. The
expectation value of the second term in vanishes because the signal
does not correlate with the noise. Finally, the expectation value of
the third term is given by
\begin{eqnarray}
   \langle \eta^T Q \, \eta \rangle 
    & = & \sum_{i,j=1}^m Q_{ij} \int_0^T \langle \eta_i(t - \tau_i) 
          \eta_j(t - \tau_j) \rangle \, 
          dt \nonumber \\
    & = & \sum_{i,j=1}^m Q_{ij} \int_0^T \delta_{ij} 
          \langle \eta_i^2 (t) \rangle \, dt 
      =   \sigma^2 \, \tr (Q) ,
\end{eqnarray}
where we assumed, for simplicity, that the noise in the detectors is 
Gaussian and its variance is $\sigma$. By transforming $Q$ to the 
diagonal form, one can show that $\tr (Q) = m\! -\! 2$. Therefore,
\begin{equation}
   \langle R(\phi, \theta) \rangle = 
      \| Q A_s h_s \|^2 + \sigma^2 \, (m-2) .
\end{equation}
At the true location of the source $Q A_s = 0$, and the function 
$\langle R(\phi, \theta) \rangle $ reaches its minimum. This concludes 
the solution to the second part of the inverse problem: determination 
of the source position (see figure~\ref{resultsHLG}). It is important
to note that the function  
$\langle R(\phi, \theta) \rangle$ may have several minima on the sky 
of which only one corresponds to the true location of the source.

\section{Examples of the inverse problem}

The simplest types of the inverse problem occur in networks of 2 and
3 detectors.

\subsection{$m = 2$}
\label{m=2inverseP}

In 2 dimensions the vectors ${\mathbf{F}}_{+}$ and
${\mathbf{F}}_{\times}$ span the entire vector space of columns of
$A$, and there are no null vectors. In this case the inverse problem
allows an exact solution 
\begin{equation}
   h = A^{-1} \, x ,
\end{equation}
for which the residual vanishes identically. It is worthwhile to
obtain this result in a somewhat different way. Note that for $m=2$
the matrix $A$ is square and therefore $A^{\dagger} = A^{-1}$. Then 
\begin{equation}
   P = A A^{\dagger} = I , \qquad {\mathrm{and}} \qquad Q = 0 .
\end{equation}
Consequently, $R(\phi, \theta) = 0$ for every point on the sky, and 
no source localization is possible. In this case, every point in the 
sky, including the one which corresponds to the true source, can be 
viewed as a minimum of $\langle R(\phi, \theta) \rangle$.

\subsection{$m = 3$}

In 3 dimensions the vectors ${\mathbf{F}}_{+}$ and
${\mathbf{F}}_{\times}$ span a 2-dimensional subspace of the 
vector space of columns of $A$. The complementary subspace, which is 
the space of null vectors, is therefore 1 dimensional, and is defined 
by one null vector, ${\mathbf{K}}$. The null condition $A^T K = 0$ can
be written in vector notation as 
\begin{equation}
   {\mathbf{K}} \cdot {\mathbf{F}}_{+} = 
   {\mathbf{K}} \cdot {\mathbf{F}}_{\times} = 0 ,
\end{equation}
which implies that up to a multiplicative constant
\begin{equation}\label{nullVectK}
   {\mathbf{K}} = {\mathbf{F}}_{+} {\mathbf{\times}} \, 
      {\mathbf{F}}_{\times} .
\end{equation}
Knowing that $Q$ projects onto the null space, we obtain 
\begin{equation}
   Q_{ij} = \frac{K_i K_j}{|{\mathbf{K}}|^2} , 
      \qquad {\mathrm{and}} \qquad
   P_{ij} = \delta_{ij} - \frac{K_i K_j}{|{\mathbf{K}}|^2} .
\end{equation}
Then from the definition (\ref{defResidual}) we find the 
residual functional,
\begin{equation}
   R \equiv x^T Q \, x = 
      \frac{\|{\mathbf{K}} \cdot {\mathbf{x}}\|^2}{|{\mathbf{K}}|^2},
\end{equation}
which is the minimization functional of G\"{u}rsel and Tinto.

Next, we introduce two vectors to partition the Moore-Penrose inverse:
\begin{equation}
   A^{\dagger} = [ \, {\mathbf{H}}_{+} \; {\mathbf{H}}_{\times} ]^T ,
\end{equation}
so that the solution for $h$ (\ref{LSQsolution}) can be written as 
\begin{equation}\label{solutionGT}
   h_{+} = {\mathbf{H}}_{+} \cdot {\mathbf{x}} , 
      \qquad {\mathrm{and}} \qquad
   h_{\times} = {\mathbf{H}}_{\times} \cdot {\mathbf{x}} .
\end{equation}
It is easy to show that the partition vectors are given by
\begin{eqnarray}
   {\mathbf{H}}_{+}  & = & \frac{1}{|{\mathbf{K}}|^2} \; 
      {\mathbf{F}}_{\times} \times {\mathbf{K}} , \label{defHplus} \\
   {\mathbf{H}}_{\times} & = & \frac{-1}{|{\mathbf{K}}|^2} \; 
      {\mathbf{F}}_{+} \times {\mathbf{K}} . \label{defHcross} 
\end{eqnarray}
Indeed, from the definition of the Moore-Penrose inverse
(\ref{MPinverse}) we find that
\begin{eqnarray}
   {\mathbf{H}}_{+} & = & 
      [M^{-1}]_{11} \, {\mathbf{F}}_{+} +
      [M^{-1}]_{12} \, {\mathbf{F}}_{\times}  \nonumber \\
                    & = & (\det M)^{-1} \; \left[ 
      ({\mathbf{F}}_{\times} \cdot {\mathbf{F}}_{\times}) \,
       {\mathbf{F}}_{+} - 
      ({\mathbf{F}}_{+}      \cdot {\mathbf{F}}_{\times}) \, 
       {\mathbf{F}}_{\times} \right] \nonumber \\
                    & = & (\det M)^{-1} \;\; 
      {\mathbf{F}}_{\times} \times 
     ({\mathbf{F}}_{+} \times {\mathbf{F}}_{\times}) .
      \label{derivHplus}
\end{eqnarray}
Combining this result with 
\begin{equation}
   \det M = |{\mathbf{K}}|^2 ,
\end{equation}
we obtain (\ref{defHplus}). Similarly, one can derive (\ref{defHcross}).
Note that the solution for $h$, written in terms of ${\mathbf{H}}_{+}$ 
and ${\mathbf{H}}_{\times}$, is the waveform estimator of
G\"{u}rsel and Tinto.

\section{Difficulties with the LSQ minimization}

Direct minimization of the LSQ functional encounters various
difficulties, some of which are briefly described in this section.

\subsection{Divergence of the expectation value of the solution}
\label{divergeMean}

Consider the solution given by the Moore-Penrose inverse
(\ref{LSQsolution}) as a function of sky position: 
\begin{equation}
   h(t) = M^{-1}(\phi, \theta) A^T(\phi, \theta) \, x(t) .
\end{equation}
As we have seen, on average, this solution correctly reproduces the 
waveform of the gravitational wave, provided that the estimated
location of the source coincides with its true location. If, however,
the estimated source location slightly deviates from the true
location, the solution can be very different from the true waveform,
especially near those places on the sky where the matrix $M$ becomes
singular. Further discussion of the variations of the estimated
waveforms and solutions based on the penalty functional can be found
in \cite{Mohanty:2006}.

\subsection{Divergence of the variance of the solution}
\label{divergeVar}

Consider now the error in estimation of the gravitational-wave
amplitudes which comes from the presence of noise in the data:
\begin{equation}
   \delta h(t) = A^{\dagger}(\phi, \theta) \eta(t) .
\end{equation}
Taking into account the fact that the noise in the detectors is
uncorrelated, we obtain
\begin{eqnarray}
   \langle \delta h_i(t) \, \delta h_j(t') \rangle
      & = &  \sum_{k,l=1}^m [A^{\dagger}]_{ik} [A^{\dagger}]_{jl} 
             \langle \eta_k(t) \, \eta_l(t') \rangle \nonumber \\
      & = &  \sigma^2 [M^{-1}]_{ij} \delta(t - t') .
\end{eqnarray}
Therefore, the variance of the solution,
\begin{equation}
   \int_0^T \langle \delta h_i^2(t) \rangle \, dt = 
      \sigma^2 [M^{-1}]_{ii} ,
\end{equation}
diverges as the estimated source location approaches those places 
in the sky where the matrix $M$ becomes singular. Solutions to this
problem based on constraints applied to the waveforms can be found 
in \cite{Klimenko:2005}.

\subsection{Divergence of the residual functional}
\label{discRes}

Finally, consider the residual functional (\ref{defResidual}). 
According to this definition the residual diverges if matrix $M$
becomes singular. This result becomes somewhat puzzling if we recall
that in one particular case of singularity, namely when all detectors
in the network are co-aligned, the residual in fact is well defined. 
Indeed, consider a network of $m$ co-aligned detectors. In this case
it is not possible to solve for both $h_+$ and $h_{\times}$, and one 
can only solve for their linear combination: 
$\xi = F_{+} h_{+} + F_{\times} h_{\times}$. Then the LSQ 
functional takes the form \cite{Anderson:2001}:
\begin{equation}\label{LSQdef2}
   L[\xi] = \sum_{i=1}^m \| x_i(t) - \xi(t) \|^2 ,
\end{equation}
with an obvious solution: $\xi = \frac{1}{m} \sum_{i=1}^m x_i$. 
In this case, the projectors $P$ and $Q$ can be defined as
\begin{equation}\label{coalignDetPQ}
   P_{ij} = \frac{1}{m} , 
      \qquad {\mathrm{and}} \qquad 
   Q_{ij} = \delta_{ij} - \frac{1}{m} .
\end{equation}
Consequently, the residual functional is given by
\begin{equation}\label{coalignR}
   R \equiv x^T Q \, x = \frac{1}{m} \sum_{i,j=1}^m
      \int_0^T [x_i^2(t) - x_i(t) x_j(t)] \, dt .
\end{equation}
Mathematically, this is an altogether different inverse problem, 
and there is no obvious reason why the solution of (\ref{LSQdef2}) 
must be related to the solution of the original inverse problem 
(\ref{LSQdef}). However, our physical intuition tells us that the
network of nearly aligned detectors must behave very similarly to a
network in which the same detectors become fully aligned. Yet, the
residual calculated for nearly aligned detectors (\ref{defResidual}) 
does not approach the one calculated for co-aligned detectors
(\ref{coalignR}) in the limit when the detectors become perfectly
aligned. This discontinuity of the residual functional becomes even
more striking in two-detector networks for which $R = 0$ (see 
section~\ref{m=2inverseP}) for all detector orientations, no matter 
how close to perfect alignment they are. In this case the discrepancy 
was called the {\it two-detector paradox} 
\cite{Johnston:2004, Klimenko:2005, Mohanty:2006}.

\section{Rank deficiency}

All the above problems originate from rank deficiency of the network
response matrix $A$. Recall that the nominal rank of $A$ is 2. The
rank of $A$ drops to 1 if all the rows of $A$ become proportional to
each other, which is known as full row degeneracy. Column degeneracy 
occurs when the two columns of matrix $A$ become proportional to each 
other. Note that full row degeneracy implies column degeneracy of $A$ 
and vice versa. In other words, either degeneracy results in
collinearity of the range vectors:
\begin{equation}\label{FpropF}
   {\mathbf{F}}_{\times} = \beta \, {\mathbf{F}}_{+} ,
\end{equation}
where $\beta$ is real. If this condition is satisfied, the rank of 
$A$ is 1. The simplest case of rank deficiency occurs when all 
detectors in the network are co-aligned and therefore all the rows
of matrix $A$ are equal.

\begin{figure}[t]
   \includegraphics[width=\textwidth]{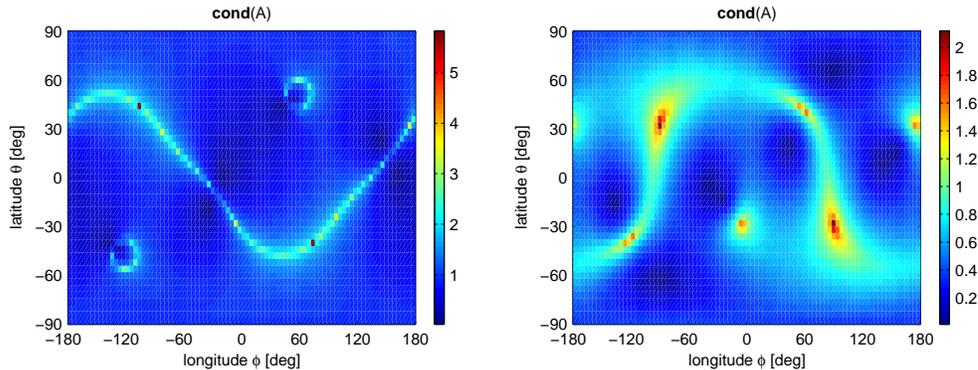}
   \caption{Condition number as a function of sky position 
   (colormap in log scale). {\it Left:} LIGO-only network (H1-H2-L),
   {\it Right:} LIGO-GEO network (H1-H2-L-G). Note that close
   alignment of LIGO interferometers leads to very high condition
   number for some locations on the sky ($\gtrsim 10^5$), and the
   inclusion of GEO detector results in significant reduction of 
   the condition number ($\gtrsim 10^2$).}
   \label{cond2net}
\end{figure}

In practice, we are seldom concerned with the exact collinearity of
$\mathbf{F}_{+}$ and $\mathbf{F}_{\times}$ as the problem already
occurs when these vectors are close to being collinear. This can be
seen from the fact that
\begin{equation}
   \det M = |{\mathbf{F}}_{+}|^2 |{\mathbf{F}}_{\times}|^2 - 
      ({\mathbf{F}}_{+} \cdot {\mathbf{F}}_{\times})^2 
      \rightarrow 0 ,
\end{equation}
as the two range vectors approach collinearity, making the
Moore-Penrose inverse divergent. Quantitatively, the 
degree to which the inversion of $A$ becomes ill defined is described
by the {\it condition number} \cite{Hansen:1998}:
\begin{equation}
   {\mathbf{cond}}(A) = \| A \| \cdot \| A^{\dagger} \| ,
\end{equation}
where $\| . \|$ stands for matrix 2-norm.\footnote{The 2-norm of 
matrix $A$ is defined as the maximum value of the 2-norm of vector
$A\mathbf{v}$ under condition that $\| \mathbf{v} \|=1$.}
Perfectly-invertible orthogonal matrices have condition number 
of 1. Large condition numbers indicate ill-defined inversion.
Figure~\ref{cond2net} shows condition numbers for two examples 
of detector networks. The locations in the sky with large condition 
number correspond to rank-deficient network response matrix.

\section{Tikhonov regularization}

Several techniques are available in applied mathematics to address
rank deficiency of the coefficient matrix in the LSQ problem 
\cite{Hansen:1998}. One of the most commonly used and best understood
techniques is the Tikhonov regularization method \cite{Tikhonov:1977}.
The key idea in this method is to introduce a regularization
functional (regulator) $\Omega[h(t)]$ with parameter $g>0$ such that 
the modified LSQ functional,
\begin{equation}\label{TikhLSQ}
   L_g[h] = \| x(t) - A \, h(t) \|^2 + g \; \Omega[h] ,
\end{equation}
no longer suffers from rank deficiency. Consider for example a
quadratic regulator,
\begin{equation}
   \Omega[h] \equiv h^T \Omega \, h = \sum_{i,j=1}^m
      \int_0^T \Omega_{ij} h_i(t) h_j(t) \, dt ,
\end{equation}
where $\Omega_{ij}$ is a symmetric 2$\times$2 matrix. Quadratic
regulators preserve the linearity of the inverse problem and therefore
have an advantage over other forms. Then the solution of the inverse
problem becomes
\begin{equation}\label{TikhSol}
   h = M_g^{-1} A^T x , \qquad {\mathrm{where}} \qquad 
   M_g = M + g \, \Omega .
\end{equation}
Therefore, the regularized version of the Moore-Penrose inverse is
\begin{equation}
   A_g^{\dagger} = M_g^{-1} A^T ,
\end{equation}
which leads to the following generalization of matrices $P$ and $Q$:
\begin{equation}
   P_g = A A_g^{\dagger} ,
   \qquad {\mathrm{and}} \qquad 
   Q_g = I - A A_g^{\dagger} .
\end{equation}
In general, these matrices no longer satisfy the property of
projectors. However, the connection between the matrix $Q$ and the
residual functional still holds. Indeed, substituting the solution 
(\ref{TikhSol}) into the modified LSQ functional (\ref{TikhLSQ}), we 
obtain 
\begin{equation}\label{residualRg}
   R_g = x^T Q_g \, x ,
\end{equation}
which is equivalent to (\ref{defResidual}) despite the presence of
a regulator in $L_g[h]$.

The role of $\Omega$ is to render the inverse of $M_g$ well defined
when $M$ is nearly singular. This resolves the problems associated
with singularities of $M$, described in sections \ref{divergeMean} 
and \ref{divergeVar}. Introduction of the regulator also solves the 
problem of discontinuity of the residual, described in
section~\ref{discRes}. Indeed, for $g \neq 0$ the residual $R_g$ is 
a continuous function of the detector alignment, and no divergence 
occurs in $R_g$ when the matrix $A$ becomes rank deficient 
(\ref{FpropF}). Furthermore, explicit calculations show that in this
case
\begin{equation}
   [P_g]_{ij} = \frac{\alpha}{\alpha + g \det \Omega } \; f_i f_j , 
\end{equation}
where $\mathbf{f}$ is a unit vector along $\mathbf{F_{+}}$, and 
\begin{equation}
   \alpha = (\beta^2 \Omega_{11} - 2 \beta \Omega_{12} + \Omega_{22}) 
   |\mathbf{F_{+}}|^2 .
\end{equation}
If the detectors in the network become co-aligned,  
$f_i = \frac{1}{\sqrt{m}}$ and therefore 
\begin{equation}
   [P_g]_{ij} = \frac{\alpha}{\alpha + g \det \Omega } \; \frac{1}{m} .
\end{equation}
One can see from this expression that $P_g$ reduces to $P$ in 
(\ref{coalignDetPQ}), in the limit $g \rightarrow 0$, and
consequently, $Q_g$ reduces to $Q$ in (\ref{coalignDetPQ}). Hence, 
the residual functional (\ref{residualRg}) reduces to that in
(\ref{coalignR}).

\section{Optimum condition regulator}

As we have seen, the degree to which matrix $A$ becomes non-invertible,
measured by the condition number, strongly depends on the sky location. 
It is therefore desirable to construct a regulator which is a function 
of sky position. Particularly useful will be regulators which adjust 
themselves to higher condition number, always guaranteeing well-defined 
inversion of $A$. Here we construct one example of such a regulator.

\begin{figure}[t]
   \includegraphics[width=\textwidth]{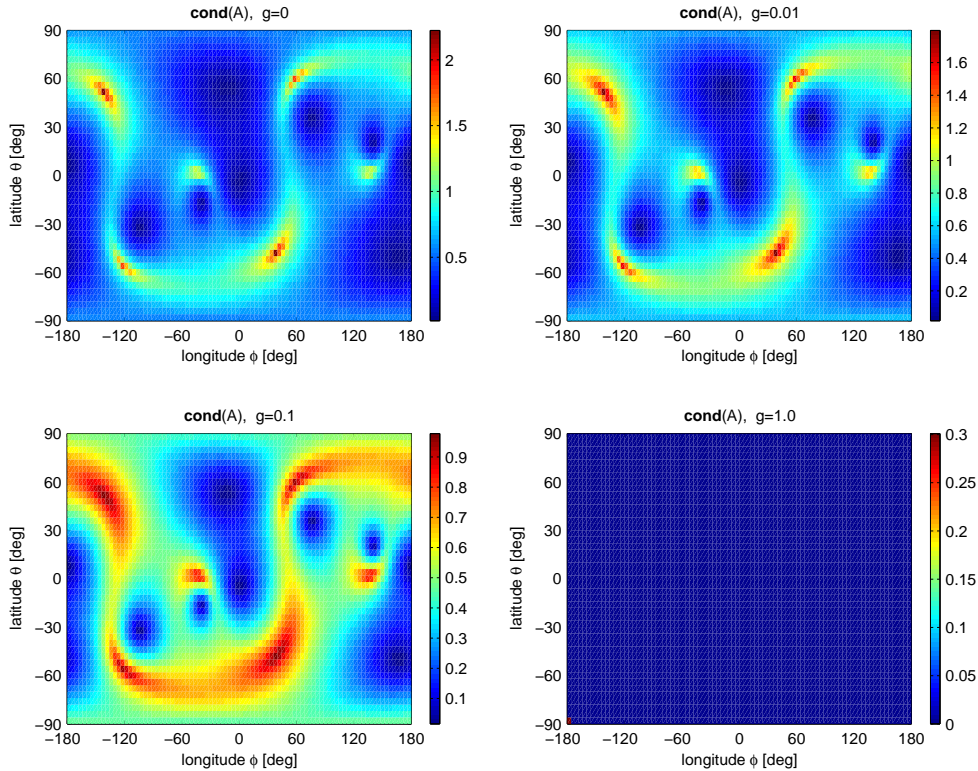}
   \caption{Condition number for LIGO-VIRGO network (H1-H2-L-V) as
   a function of sky position for different values of $g$ (colormap in 
   log scale). Complete regularization takes place for $g=1$ 
   (bottom-right) in which case the condition number is 1.}
   \label{condOparam}
\end{figure}

Consider matrix $M$ in the space of its eigenvectors:
\begin{equation}
   \tilde{M}_{ij} = \left[
   \begin{array}{cc}
   \mu_1  &     0      \\
     0    &  \mu_2
   \end{array}\right] ,
\end{equation}
It is easy to show that its eigenvalues, $\mu_1$ and $\mu_2$, are
always positive and one is always greater than the other, e.g. 
$\mu_1 > \mu_2$. Since the purpose of a regulator is to prevent 
singularities which occur when $\mu_{1,2} \rightarrow 0$, it is
sufficient to consider $\Omega$ which is diagonal in the space of
eigenvectors of $M$:
\begin{equation}
   \tilde{\Omega}_{ij} = \left[
   \begin{array}{cc}
   \omega_1  &     0      \\
        0    &  \omega_2
   \end{array}\right] ,
\end{equation}
where we assume that $\omega_{1,2} \geq 0$ so that $\Omega[h]$ is 
positive definite.

\begin{figure}[t!]
   \includegraphics[width=\textwidth]{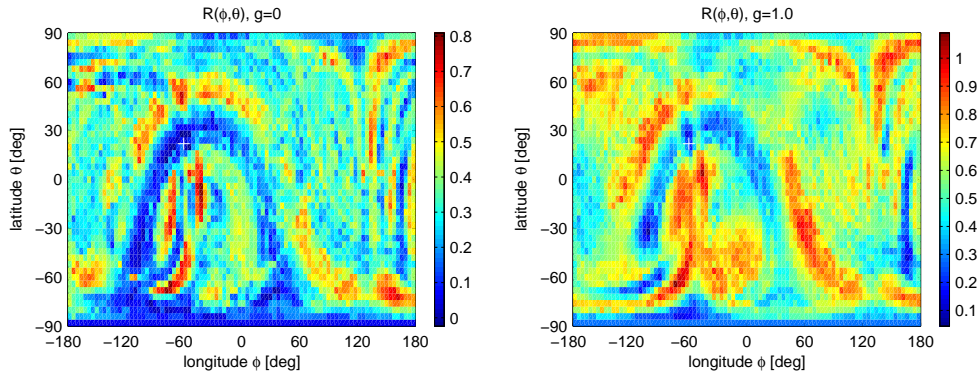}
   \caption{The residual functional (\ref{residualRg}), for 
   LIGO-VIRGO network (H1-H2-L-V) with no regularization (left) and 
   full regularization (right). The injected waveforms (figure 1) 
   are amplified to produce the signals in detectors with the SNR of 
   8, 8, 12, and 10 for source location $(-60^{\circ}, 22^{\circ})$ 
   (marked by $+$). Note a significant reduction in the degeneracy 
   (blue area) introduced by the regulator.}
   \label{residOparam}
\end{figure}

In most cases of interest only one of the eigenvalues, $\mu_2$, can 
be singular. If both eigenvalues vanish then
$\mathrm{tr}(M)=0$. This would imply that $|\mathbf{F}_{+}|^2 + 
|\mathbf{F}_{\times}|^2 = 0$, which, in turn, implies that all
components of vectors $\mathbf{F}_{+}$ and $\mathbf{F}_{\times}$
vanish. In a given detector both antenna-pattern functions vanish 
only if the source is located in the plane of the detector, on the 
bisector of the detector arms, or on the normal to the bisector. 
Apart from the two Hanford interferometers, a general network of 
detectors does not have common arm bisectors or bisector normals, 
which is why both eigenvalues cannot vanish simultaneously.

A singularity in which one of the eigenvalues approaches
zero whereas the other remains finite is known as a {\it finite gap}
\cite{Hansen:1998}. In this case, it is sufficient to regularize only
the smallest eigenvalue $\mu_2$. We therefore limit ourselves to 
regulators in which $\omega_1 = 0$, and which are often called 
semi-norms in the space of solutions. Note that the parameter $g$ 
becomes redundant; it can be absorbed into $\omega_2$. Nevertheless, 
we retain $g$ so that we can control the strength of the regulator, 
and assume that the maximum value of $g$ is 1. We can now construct 
a regulator which makes the matrix $A$ fully invertible over the entire
sky. Quantitatively, this means that for $g=1$
\begin{equation}
   \mathbf{cond}(A) = \| A \| \cdot \| A_g^{\dagger} \| = 1 ,
\end{equation}
for all $\phi$ and $\theta$. Taking into account that 
\begin{equation}
   \| A \| = \sqrt{\mu_1}, \qquad {\mathrm{and}} \qquad 
   \| A_g^{\dagger} \| = \mathrm{max} \left( \frac{1}{\sqrt{\mu_1}}, 
      \frac{\sqrt{\mu_2}}{\mu_2 + \omega_2} \right) ,
\end{equation}
we find that the minimum value for such $\omega_2$ is given by 
\begin{equation}
   \omega_2^* = \sqrt{\mu_1 \mu_2} - \mu_2 .
\end{equation}
In other words, any regulator of the semi-norm type in which 
$\omega_2 > \omega_2^*$ yields the inversion of $A$ with condition 
number of 1.

Figure \ref{condOparam} shows the improvement in the condition number
for the LIGO-VIRGO network which results from this regularization. This
will imply improvement in the accuracy of the solution for $h(t)$. Note 
that the introduction of the regularization also improves localization 
of the source on the sky, as shown in figure \ref{residOparam}. Further 
analysis of the role of regularization will be given elsewhere. We 
conclude with the reminder that regularization, by its nature, 
introduces a bias and therefore the optimal approach must be a
trade-off between the bias and the error due to noise.

\section{Conclusion}

We have shown that detection of gravitational-wave bursts of unknown
waveforms is a linear inverse problem which becomes ill posed because
of the rank deficiency of the underlying network response matrix. 
Following the general scheme of Tikhonov regularization, we introduced
a semi-norm regulator in the space of solutions which guarantees full 
inversion of the network response matrix over the entire sky. The 
analysis presented here is general and applies to any network of
interferometric gravitational-wave detectors. The problem of rank 
deficiency is particularly important from a practical point of view 
because the condition number for the LIGO-only network can be extremely 
high ($\gtrsim 10^5$) and regularization of the response matrix will
play a crucial role in stabilizing the solution of the inverse problem. 
For networks in which LIGO detectors are accompanied by VIRGO or GEO
interferometer the condition number is substantially better 
($\gtrsim 10^2$) and yet still in need of regularization.

\ack

I thank S.~Waldman for reminding me of the method of Tikhonov
regularization, and R.~Coldwell, L.S.~Finn, S.~Klimenko, and 
S.D.~Mohanty for valuable discussions. This work was supported by 
the Center for Gravitational Wave Physics at the Pennsylvania State 
University and the US National Science Foundation under grants 
PHY 02-44902 and PHY 03-26281. The Center for Gravitational Wave 
Physics is funded by the National Science Foundation under 
cooperative agreement PHY 01-14375. This article has been assigned 
LIGO Laboratory document number LIGO-P060005.

\section*{References}

\providecommand{\newblock}{}

\end{document}